# Molcontroller: a VMD Graphical User Interface for Manipulating Molecules

ChenChen Wu[1], Shengtang Liu[1], Shitong Zhang[1], Zaixing Yang[1*]

[1]*Institute of Quantitative Biology and Medicine, State Key Laboratory of Radiation Medicine and Protection, School of Radiation Medicine and Protection, Collaborative Innovation Center of Radiological Medicine of Jiangsu Higher Education Institutions, Soochow University, Jiangsu 215123, China.*

[*]Corresponding authors: zxyang@suda.edu.cn (Z. Yang)

***ABSTRACT:*** Visual Molecular Dynamics (VMD) is one of the most widely used molecular graphics software in the community of theoretical simulations. So far, however, it still lacks a graphical user interface (GUI) for molecular manipulations when doing some modeling tasks. For instance, translation or rotation of a selected molecule(s) or part(s) of a molecule, which are currently only can be achieved using *tcl* scripts. Here, we use *tcl* script develop a user-friendly GUI for VMD, named *Molcontroller*, which is featured by allowing users to quickly and conveniently perform various molecular manipulations. This GUI might be helpful for improving the modeling efficiency of VMD users.

**I. INTRODUCTION**

Visual Molecular Dynamics (VMD)[1] is one of the most powerful molecular visualization and modeling program for the researchers of theoretical simulations. However, there is still a lack of a graphical user interface (GUI) for molecular manipulations in case of doing some modeling tasks, such as translation/rotation of a selected molecule(s) or part(s) of a molecule, or merging two or even more molecules into a single file, which, currently, can only be achieved using *tcl* scripts. In view of this, we develop a user-friendly GUI for VMD, named *Molcontroller*, to perform these daily used manipulations. Moreover, it can aslo provide a variety of molecular identifier information in the VMD displaying window, which should be very convenient for the molecular structure analyses. This GUI might be helpful for improving the modeling efficiency of VMD users.

**II. TECHNICAL IMPLEMENTATION and MOLCONTROLLER FEATURES**

The *Molcontroller* GUI is developed using *Tcl/Tk* toolkit, and can be integrated into the *Extensions*/*Modelling*/*Molcontroller* menu of VMD (find more details in the "Installation" guide section). This GUI contains several function boxes, for instance, *Information*, *Property*, *Manipulation* and *Save File* (Fig. 1). The major features are described as follows.

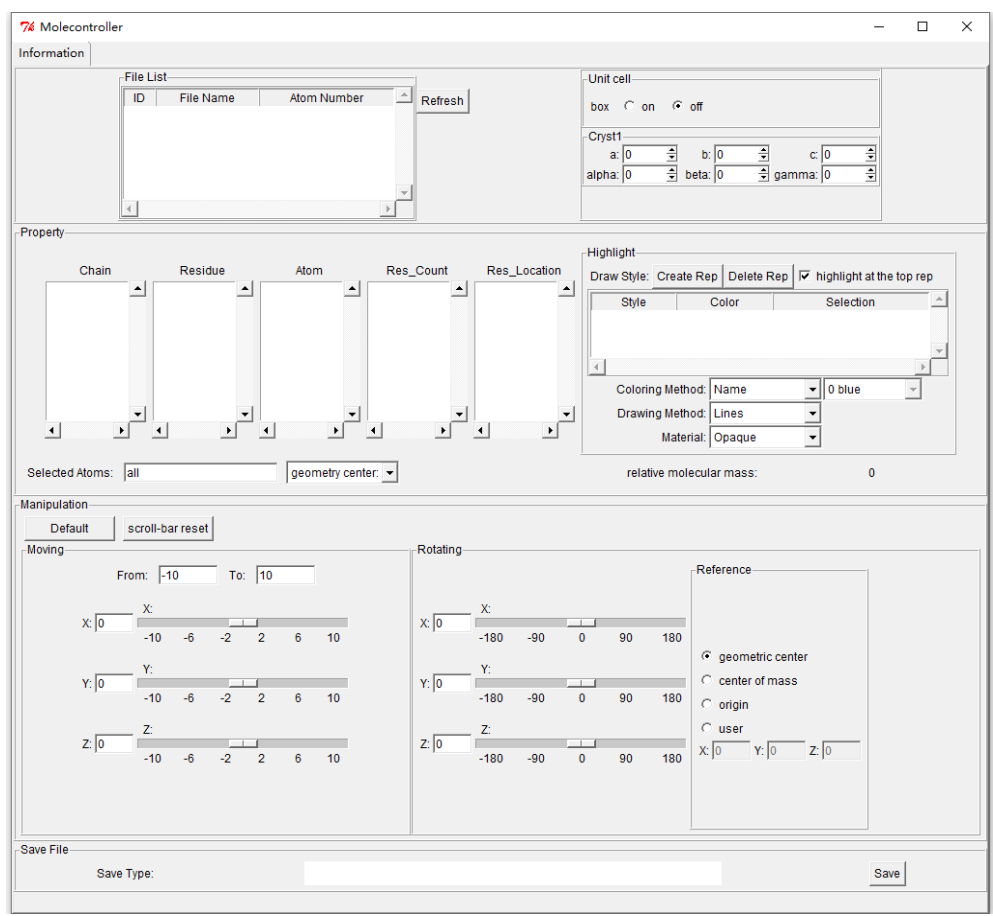

**Figure 1.** The main window of *Molcontroller* graphical user interface. It contains four function boxes: *Information*, *Property*, *Manipulation* and *Save File*.

**a) *Information***

There are two panels in the *Information* box. In the left panel (*File List*), the targeting molecule(s) for modeling manipulations is listed. In the right panel (*Unit Cell*), the user can switch the simulation box of the selected molecule on or off using the radio button. Moreover, the user can also directly modify the VMD unit-cell parameters, i.e., *a*, *b*, *c*, *alpha*, *beta* and *gamma*.

**b) *Property***

The *Property* box can display various molecular identifier information. The first three columns of the *Property* box, from *Chain* >> *Residue* >> *Atom*, display the molecular information in a tree-based structure. The column of *Res_Count* displays the total number of each type of residue in a selected molecule. If a specific residue is clicked in the column of *Res_Count*, all residual IDs of the same residue will be listed in the column of *Res_Location*. The user can click the selections in the *Property* box to display the entire or some specific parts of a molecule. Alternatively, this can also be achieved by typing the text commands of VMD in the *Selected Atoms* text entry. In the box of *Highlight Method*, the user can render the selected molecule (or its part(s)) using the routine drawing style and coloring method of VMD software on the topmost representation with click *highlight at the top rep* checkbox. In addition, the geometric center/the center of mass (COM) of the molecule and relative molecular mass of the selected molecule (or its part) are displayed at the bottom of *Information* box.

**c)** *Manipulation*

The main function of *Manipulation* box is to implement translation or rotation of a selected molecule (or its part(s)). The moving range can be defined in the text entry boxes labeled with *From* and *To*. The selected molecule(s) or selection(s) of a molecule can be translated and rotated along the specified *x*, *y* or *z*-axis with preset values that declared in the *Moving* and *Rotating* text entry boxes, respectively. Alternatively, these two operations can also be done through dragging the specific slider(s) along the *x*, *y* or *z*-axis. The rotation reference point of the molecule can be set to the geometric center or the COM of the molecule, the origin of the system, or a user-defined reference point using the radio button. The *Scroll bar reset* button can reset the moving slider to the initial value for further adjustments. The *Default* button can restore the original structure of the molecule for the user to readjust its structure.

**c)** *Save File*

The last box of the GUI is designed to save the current molecule or selection(s) of a molecule after moving or rotating operation(s). The user can also merge these processed molecules into a single PDB file as a complex. The file-types for saving are available in the drop-down menu at the bottom of the *Save File* box.

**III. SPECIFIC EXAMPLES**

### a) *Molecular Manipulations*

To show the molecular manipulations feature of our GUI, we simply construct an antiparallel β-sheet consisting of two Aβ$_{16-21}$ peptides (KLVFFA), which is the core amyloidogenic segment of amyloid fibril in Alzheimer's disease. The ideal targeting configuration for constructing is the KLVFFA β-sheet crystal structure that has been deposited in the protein data bank with the PDB ID of 3OW9[2] (Fig. 2a). All the translation and rotation parameters during the manipulation process are derived from the KLVFFA β-sheet crystal structure. First, two identical KLVFFA peptides are loaded into VMD. Then, opening the *Molcontroller* GUI and clicking the *Refresh* button in the *Information* panel, the two peptides are loaded in the GUI. Meanwhile, their file IDs, file names, and atom numbers are also listed in the *File List* box. Obviously, the initial coordinate of the two peptides is completely overlapped (Fig. 2b), which means that to build an antiparallel β-sheet structure, one of the peptides needs to be translated and rotated accordingly. Hence, a peptide is selected and translated 4.79 Å along with the negative direction of *y*-axis (Fig. 2c). Then, the COM of this peptide is assigned as the rotation reference point, and the peptide is rotated 180º along the *y*-axis (Fig. 2d). Furthermore, it is translated 1.02 Å along *x*-axis (Fig. 2e). Finally, the two peptides are merged into a single PDB file completing the manipulation task (Fig. 2f).

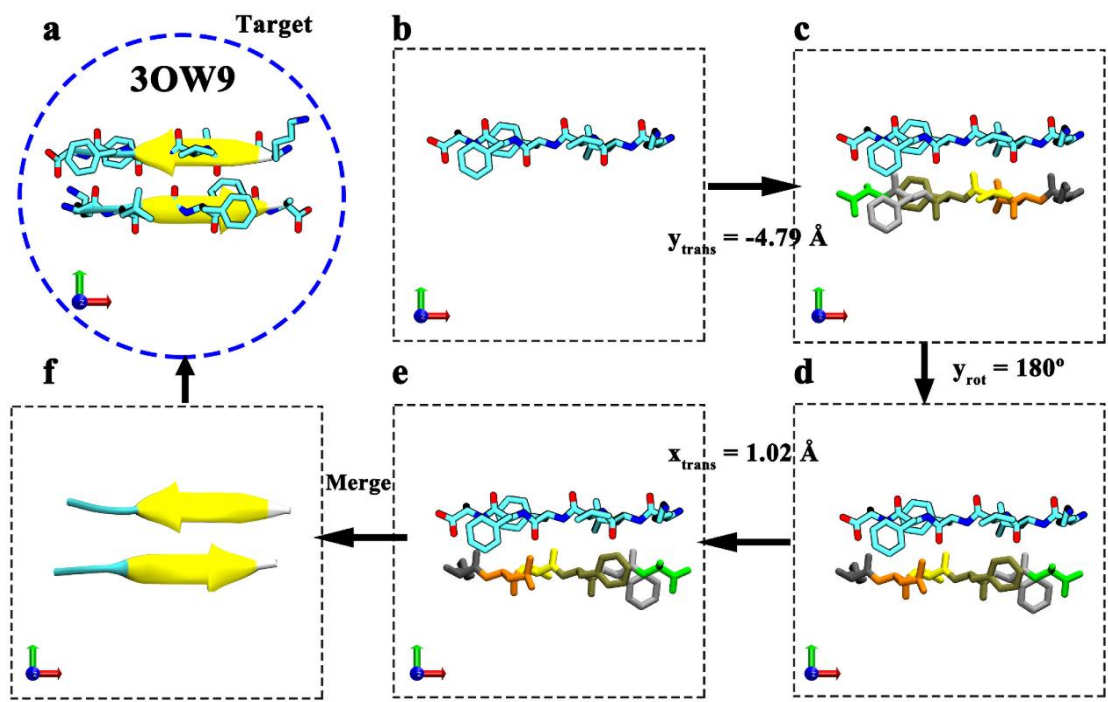

Figure 2. (a) The targeting configuration of the KLVFFA dimeric antiparallel β-sheet (PDB ID: 3OW9) shown as the *Newcartoon* model and colored with the secondary structure. (b−f) The schematic of manipulation process of building KLVFFA dimeric anti-parallel β-sheet starting with two identical KLVFFA peptides. All the translation and rotation parameters are obtained from the crystal structure (PDB ID: 3OW9).

**b)** *Structural analysis*

To show the feature of molecular structure analysis of this GUI, we use a typical potassium-channel protein, KcsA (PDB ID: 2K1E)[3], as the example (Fig. 3). Firstly, the PDB file of KcsA is loaded through molecule file browser. Then, when opening the *Molcontroller* GUI and clicking the *Refresh* button in the *Information* panel, users can find that the file ID, file name, and atom number of the protein in the PDB file are listed in the *File List* box. Meanwhile, in the property box, many molecular identifier information can be available. For example, in the *chain* column, it can be clearly seen that KcsA contains four chains, form chain A to chain D. When clicking chain A, one can find that all residues in chain A are listed in the *Residue* column. When users click any residue in chain A in the *Residue* column, all atoms of the selected residue and the total number of each type of residue will be listed in *Atom* column and *Res_count* column, respectively (Fig. 3a). When an exact type of residue is selected, all residual IDs of this type of residue will be listed in *Res_Location* column. Here, chain A, chain B, chain C, and chain D are shown with *Newcartoon* style and colored with orange, yellow, cyan, and tan, respectively. In addition, all Arg residues are selected and highlighted with van der Waals (vdW) ball style for the subsequent structural analysis (Fig.3 b). Clearly, plentiful molecular identifier information integrated into this GUI endow researches more convenience when they perform analyzing tasks on some protein structures.

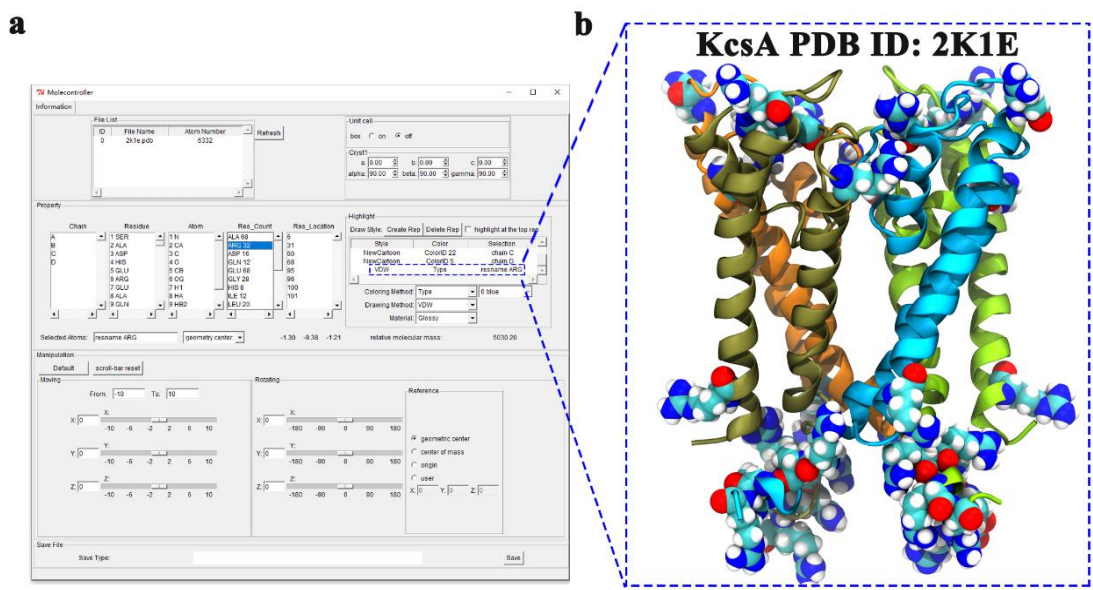

Figure 3. (a) Structural analysis of KcsA protein using *Molcontroller*. (b) Chain A, chain B, chain C, and chain D are drawn with *Newcartoon* style and colored with orange, yellow, cyan, and tan, respectively. All 32 arginine residues are highlighted with vdW balls.

## IV. INSTALLATION

The plugin can be added into the *Extensions* submenu in the VMD menu bar. To install *Molcontroller* on Windows, Linux or MacOS, follow the following steps:

1. Create a personal plugin directory, recommended as follows:

VMD-Home/plugins/noarch/tcl/molcontroller1.0, where VMD-Home is VMD installation directory.

2. Download the scripts; archive and extract it in the above directory:

The source code is released under the GNU[4] public license version 3 from https://github.com/ChenchenWu-hub/Molcontroller.

3. Add the following line to your configuration file of VMD (e.g., VMD- Installation-Directory/vmd/vmd.rc for Windows, VMD- Home/lib/vmd/.vmdrc for Linux, and VMD-Home/Contents/vmd/.vmdrc for MacOS, respectively):

lappend auto_path VMD- Installation-Directory/plugins/noarch/tcl/molcontroller1.0

vmd_install_extension molcontrol molcontroller_tk "Modeling/Molcontrol"

If the plugin installation is successful, you will see the plugin under *Extensions/Modelling/Molcontroller* in the VMD menu.

**CONCLUSION**

We develop a user-friendly GUI for the molecular graphics software VMD, which allows researchers of theoretical simulations to conveniently translate and rotate molecule in the modeling process. This GUI is an open-source program, which can be obtained freely. Also, it can be freely modified to fit the specific needs of various VMD users.

**AUTHOR INFORMATION**

**Corresponding Author**

**Zaixing Yang -** *Institute of Quantitative Biology and Medicine, State Key Laboratory of Radiation Medicine and Protection, School of Radiation Medicine and Protection, Collaborative Innovation Center of Radiological Medicine of Jiangsu Higher Education Institutions, Soochow University, Jiangsu 215123, China;* orcid.org/0000-0003-3521-6867; Email: zxyang@suda.edu.cn

**Authors**

**Chenchen Wu -** *Institute of Quantitative Biology and Medicine, State Key Laboratory of Radiation Medicine and Protection, School of Radiation Medicine and Protection, Collaborative Innovation Center of Radiological Medicine of Jiangsu Higher Education Institutions, Soochow University, Jiangsu 215123, China*

**Shengtang Liu -** *Institute of Quantitative Biology and Medicine, State Key Laboratory of Radiation Medicine and Protection, School of Radiation Medicine and Protection, Collaborative Innovation Center of Radiological Medicine of Jiangsu Higher Education Institutions, Soochow University, Jiangsu 215123, China;* orcid.org/0000-0001-9632-9377;

**Shitong Zhang -** *Institute of Quantitative Biology and Medicine, State Key Laboratory of Radiation Medicine and Protection, School of Radiation Medicine and*

*Protection, Collaborative Innovation Center of Radiological Medicine of Jiangsu Higher Education Institutions, Soochow University, Jiangsu 215123, China;* orcid.org/0000-0002-7747-2036;

**Notes**

The authors declare no competing financial interests.

## ACKNOWLEDGMENTS

This work was partially supported by the National Natural Science Foundation of China (11574224 and U1967217), Collaborative Innovation Center of Radiological Medicine of Jiangsu Higher Education Institutions, Jiangsu Provincial Key Laboratory of Radiation Medicine and Protection, and A Project Funded by the Priority Academic Program Development of Jiangsu Higher Education Institutions (PAPD).